\pdfoutput=1
\documentclass [pre, reprint, amsmath, showpacs, superscriptaddress, twocolumn]{revtex4-1}

\usepackage{graphicx}
\usepackage[colorlinks=true, linkcolor=blue, citecolor=blue]{hyperref}
\usepackage{color}
\definecolor{Fixcolor}{rgb}{1.0,0.2,0.0}

\def\L{{\texttt{L}}}
\def\R{{\texttt{R}}}
\def\tot{{\mathrm{tot}}}

\begin{document}
\title{Between giant oscillations and uniform distribution of droplets --\\
       the role of varying lumen of channels in microfluidic networks}
\author{Olgierd Cybulski}
   \email{olgierd.cybulski@gmail.com}
   \affiliation{Institute of Physical Chemistry, Polish Academy of Sciences,
Kasprzaka 44/52, 01-224 Warsaw, Poland}
\author{Slawomir Jakiela}
   \affiliation{Department of Biophysics, Warsaw University of Life Sciences,
Nowoursynowska 159, 02-776 Warsaw, Poland}
   \affiliation{Institute of Physical Chemistry, Polish Academy of Sciences,
Kasprzaka 44/52, 01-224 Warsaw, Poland}
\author{Piotr Garstecki}
   \email{garst@ichf.edu.pl}
   \affiliation{Institute of Physical Chemistry, Polish Academy of Sciences,
Kasprzaka 44/52, 01-224 Warsaw, Poland}
\date{\today}

\begin{abstract}
The simplest microfluidic network (a loop) comprises two parallel
channels with a common inlet and a common outlet. Recent studies, that
assumed constant cross-section of the channels along their length, have
shown that the sequence of droplets entering left (L) or right (R) arm of
the loop can present either a uniform distribution of choices
(e.g. RLRLRL...) or long sequences of repeated choices (RRR...LLL),
with all the intermediate permutations being dynamically equivalent
and virtually equally probable to be observed.
We use experiments and computer simulations to show that even
small variation of the cross-section along channels completely
shifts the dynamics either into the strong preference for highly grouped
patterns (RRR...LLL) that generate system-size oscillations in flow,
or just the opposite -- to patterns that distribute the droplets
homogeneously between the arms of the loop. We also show the
importance of noise in the process of self-organization of the
spatio-temporal patterns of droplets. Our results provide guidelines
for rational design of systems that reproducibly produce either grouped
or homogeneous sequences of droplets flowing in microfluidic networks.
\end{abstract}
\pacs{47.55.D-, 47.60.Dx, 47.20.Ky, 47.54.-r}
\maketitle

\section{Introduction}
Here we demonstrate that \emph{variation of the cross-section of microchannels}
along their length in microfluidic networks may determine the type of
spatio-temporal patterns of droplets traveling through them. The
simplest network is a microfluidic loop with two channels having a
common inlet and a common outlet. We show that if these channels widen
downstream, the droplets are distributed homogeneously over the loop.
In contrast, narrowing of the lumen of the channels toward the
exit of the loop produces long trains of drops flowing in alternation
into each of the parallel ducts. This observation provides and important
insight and addition to the simple one-dimensional models\cite{schindler:08}
that were so far used to model the dynamics of flow of droplets in
microfluidic networks. It may also be used to construct simple
microfluidic systems that either ``homogenize'' or ``chop'' the
sequences of drops. Finally, the observation may have biological
connotations to the flow of blood in vascular networks.

Sequences of droplets flowing through -- even simple -- microfluidic
networks often produce remarkably complex and beautiful
patterns\cite{olgierdLOC,Glawdel,Iranczycy,JeanneretHamiltonian}.
This complexity may root in the interactions at many different length-scales.
Local interactions include collisions between neighboring droplets at the
T-junction\cite{BelloulCompetitionSmallLarge,CollisionPanizza,
KadivarHerminghausLoopSim} and capillary forces caused by a specific geometry
of its walls\cite{Boukellal}.
However, the most important mechanism that gives rise
to the complex dynamics, is the long-range interaction
associated with the changes in the pressure field that the droplets
both introduce -- by increasing the hydraulic resistance of
the channel they occupy -- and that the droplets respond to -- by choosing
branches with higher inflow of the continuous liquid. The general dynamics
of the flow of droplets can be analyzed with simple models of motion of
point charges of resistance along the network of one dimensional wires,
as introduced by Schindler and Ajdari\cite{schindler:08} and later used
by multiple researchers in computer simulations\cite{olgierdLOC,Glawdel,
PanizzaComplexPRL,AgentBased,Iranczycy,ParthibanFiltering,
JeanneretHamiltonian,GlawdelRobust} and analytical studies based both
directly on the discrete model\cite{Glawdel,PanizzaComplexPRL,
MaddalaVanapalliPRE} and on its continuous
generalizations\cite{AmonSelectionRules,OlgierdContinuous}

Although simplified, the 1D model captures important aspects of
collective interactions between the droplets, caused by the
modification of the hydraulic resistance of the channels in which they
flow. At diverging junctions droplets of size comparable or larger than
the cross-section of the channel can either
split\cite{TjunctionTabeling,TjunctionAfkhami,Link2004,MarioT} -- a scenario
that we do not consider here -- or enter the channel presenting momentarily
highest volumetric rate of inflow\cite{BelloulCompetitionSmallLarge}.
Since the rate of flow through a branch of a network is a function of
resistance, the inflow of a droplet into a particular microchannel influences
the trajectories of subsequent drops. The dynamics of flow of drops in networks
has been studied in detail in a spectrum of microfluidic systems, ranging from
the simplest, two-channel loops\cite{olgierdLOC,Glawdel,PanizzaComplexPRL},
long series of identical loops\cite{JeanneretHamiltonian}, successively
bifurcating cascade of loops\cite{ChoiWhitesides,AmonSelectionRules}
to a large square grid of short channels\cite{TrafficJamsBartolo}.
Even the simplest non-trivial network, i.e. the simple loop exhibits highly
complicated dynamics and complex dependencies on parameters\cite{olgierdLOC,
Glawdel,Iranczycy,PanizzaComplexPRL,AgentBased,ParthibanFiltering,
MaddalaVanapalliPRE,GlawdelRobust,AmonSelectionRules,WangVanapaliModular}
such as flow rates, intervals between droplets (or, equivalently, frequency
of feeding droplets into the system), the additional hydraulic resistance
incorporated by droplets, and length of arms of the loop.
None of these papers has considered the effects of varying cross-section of
the arms along their length on the dynamics of the system.
This factor, however, is of practical relevance, both because of the finite
precision of microfabrication that introduce undesired variation, and
due to simplicity of deliberate introduction of such a variation in
prototyping techniques.
\emph{As we show below, variation of the lumen along a fluidic branch in
a network can have critical influence on the global flow pattern.}

\begin{figure}
\centering
\includegraphics[width=0.45\textwidth]{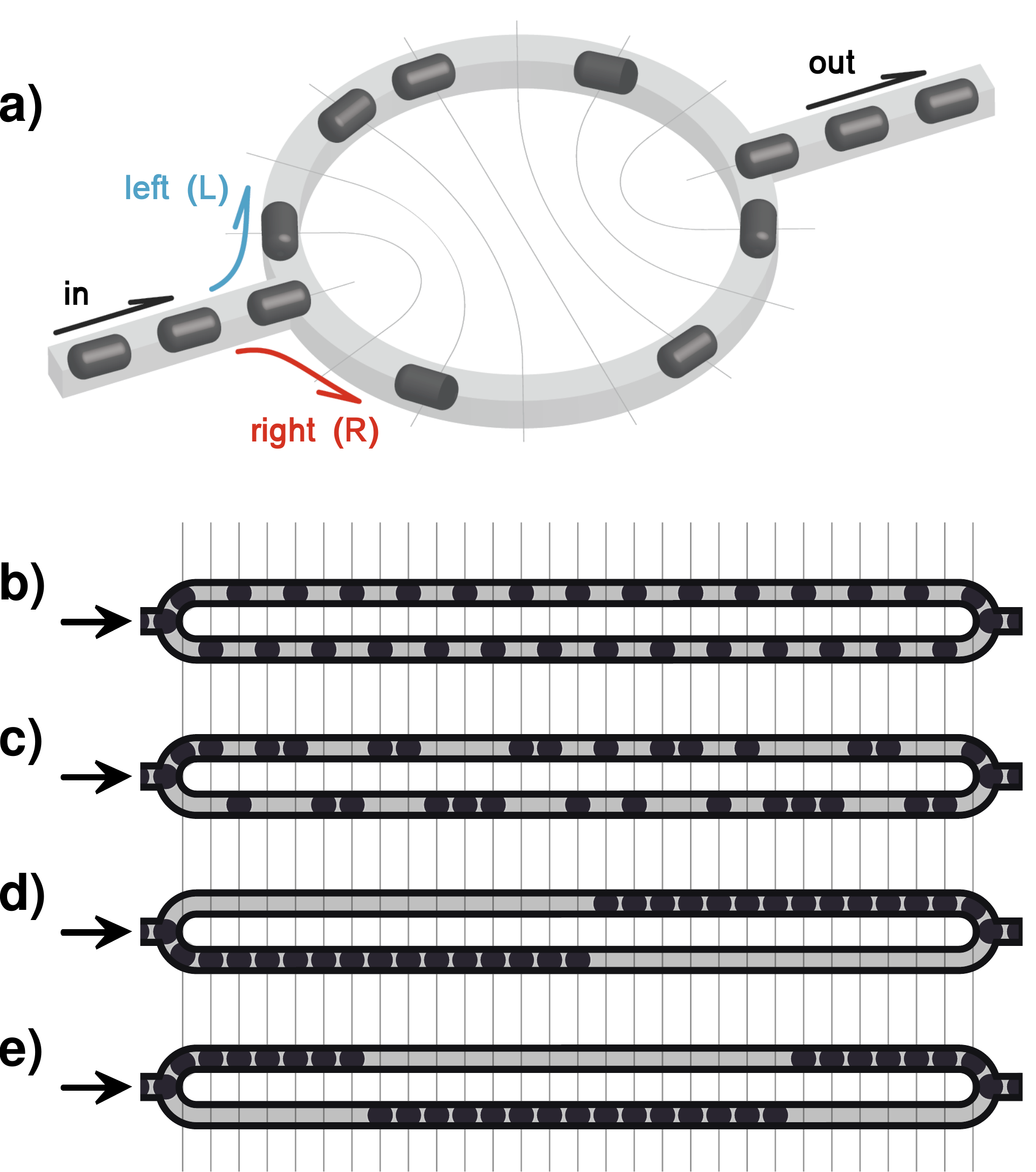}
\caption{\label{fig:im1} (color online)
(a) A schematic rendering of a symmetric
microfluidic loop. The thin gray lines mark the equi-distant points in
the parallel channels. (b-e) Exemplary momentary configurations of
droplets in the loop for three different patterns: (b) perfectly
homogeneous (chopped) sequence, (c) an intermediately chopped, random
configuration, and (d--e) perfectly queued sequence at two instants.
}
\end{figure}

Any particular droplet entering the loop (see Fig.~\ref{fig:im1}) may flow
into either the left (L) or the right (R) channel. The sequence of these
``choices'' is a good descriptor of the
dynamics of the system. This sequence can be conveniently encoded as a
binary string, composed of these two (L and R) characters. Depending on
the parameters, the additional hydraulic
resistance carried by the droplets, the volumetric flow rate and the
frequency of dripping) this string can either be periodic or not. If it
is periodic, there are many possible patterns repeated
infinitely\cite{olgierdLOC}. Which of the many different stable
states is selected, depends on the history of the system and on the
initial conditions from which the flow started. If we define $N=N_L+N_R$ as
the number of droplets in one period of the dynamics of the system, with $N_L$
and $N_R$ coding for the numbers of drops traveling through the left/right
arm, the number of distinct spatio-temporal patterns is given by
\begin{equation}
\label{eq:kombi}
\Omega\;=\;\frac{(N_L + N_R - 1)!}{N_L !\; N_R !}
\end{equation}
provided $N_L$ and $N_R$ are relatively prime numbers\cite{olgierdLOC}.
If they are not, the simple combinatorial equation overestimates the number
of patterns because it does not exclude redundant cyclic shifts. In that case
the correct evaluation of $\Omega$ can be performed using P\'olya enumeration,
as it was shown by Glawdel at al\cite{Glawdel}.

The theoretical model does not estimate the probability of falling into
any single specific pattern when starting from random initial
conditions. From computer simulations\cite{olgierdLOC} we note, that
although some patterns are easier to obtain than other ones, the
probabilities do not differ significantly. In different words, the
basins of attraction of different patterns in the space of all possible
initial conditions seem to be of comparable volume. The spectrum of all
possible patterns includes interesting limiting cases, i.e. the ones
that are maximally ``homogenized'' [Fig.~\ref{fig:im1}(b)],
and ones that are maximally ``queued''
[Fig.~\ref{fig:im1}(d) and Fig.~\ref{fig:im1}(e)]. Neither from theory nor
from simulations any of these patterns should be privileged over the vast
majority of intermediate cases (as in Fig.~\ref{fig:im1}(c)).
The probability of any of the limiting sequences to occur in a randomly
started experiment should be of the order of $1/\Omega$ -- a very small
number for large $N$ (for example, for $N=30$,  $1/\Omega < 10^{-6}$).

Thus, on the basis of the existing theory one would not expect any of
the patterns (including the chopped or queued pattern) to prevail in
the experiments. The goal of this article is to explain why this is not
necessarily true in a real experiment and how a small modification of
the system may enforce or suppress the tendency to either homogenize or
queue the distribution of droplets in networks.

\section{Experimental motivation}
Several experimental reports confirm the predictions of the simplest
model\cite{schindler:08} in systems with relatively short parallel
channels\cite{olgierdLOC,Glawdel,PanizzaComplexPRL,AmonSelectionRules,
WangVanapaliModular,MaddalaVanapalliPRE}.
The existence (i.e. the appearance and stability) of all predicted patterns
for $N_L = N_R = 4$ $(\Omega=10)$ and for smaller $N_L$, $N_R$  was
demonstrated in Ref\cite{olgierdLOC}. Adjacent bands of regular (periodic)
and irregular dynamics as well as stepwise dependence of the period on the
frequency of dripping were reported in Refs\cite{olgierdLOC,Glawdel,
PanizzaComplexPRL,WangVanapaliModular,MaddalaVanapalliPRE}.
All these results agreed with simulations and analysis based on
Schindler's model\cite{schindler:08}. In all the experiments it was
observed that after some number of repetitions the pattern may
spontaneously switch into a different one; such a behavior was
attributed to fluctuations of experimental conditions, i.e. to the
experimental noise.

\begin{figure}
\centering
\includegraphics[width=0.46\textwidth]{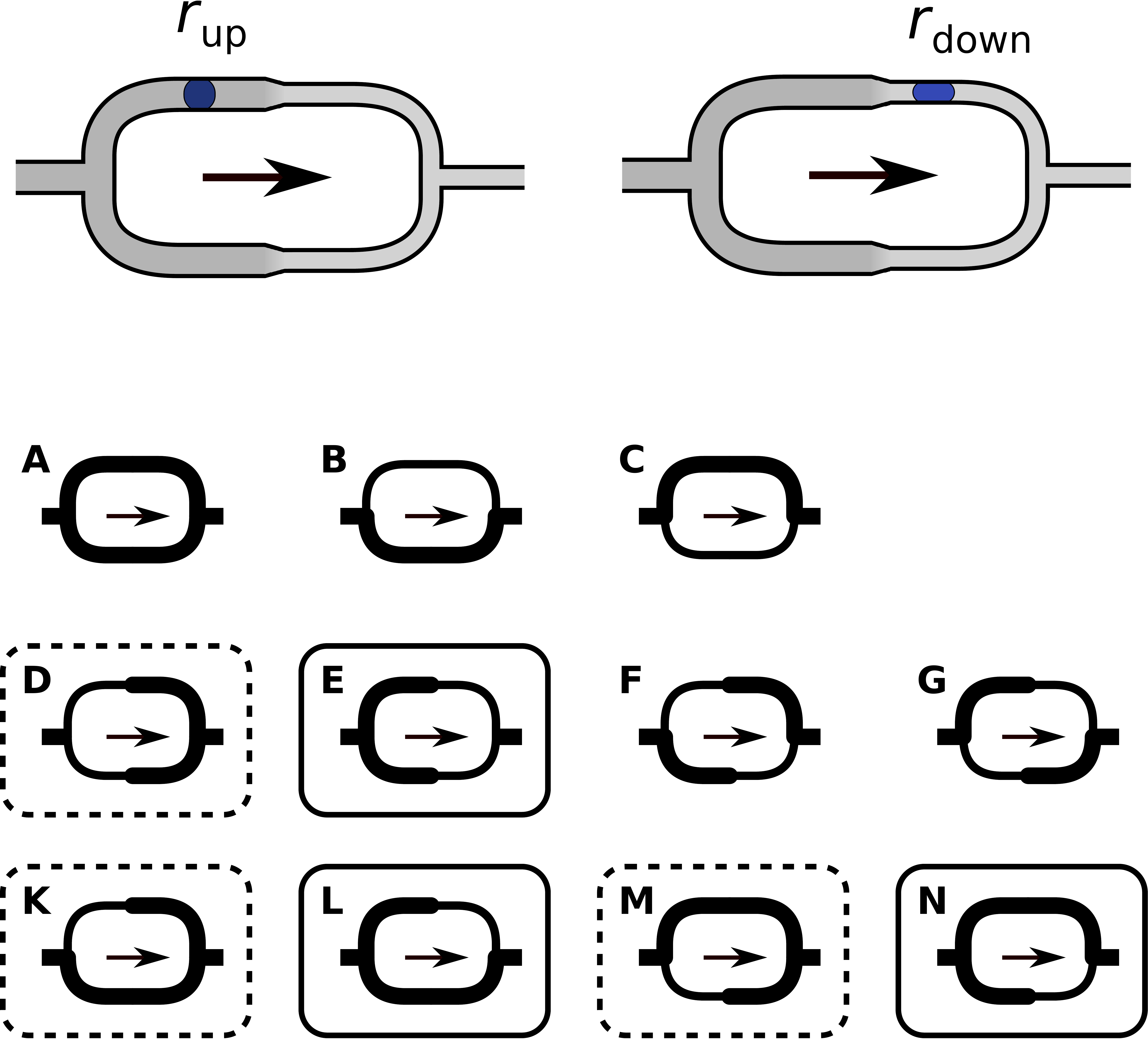}
\caption{\label{fig:im3} (color online)
Upper panel: change of the shape and additional
resistance of a droplet due to varying cross-section of channels.
Lower panel: modifications of the asymmetric loop to be tested
with respect to queuing. Thin lines correspond to the segments of
modified cross-section -- narrower than the other channels.
Configurations indicated by solid frames induced queuing, dashed frames
-- chopping (suppression of queues).
}
\end{figure}

The fundamental assumptions of the ideal model\cite{schindler:08} should
be even better fulfilled in systems comprising long channels -- i.e.
channels that can accommodate hundreds of droplets -- because
such systems are closer to the approximation of single lane ducts.
Also the role of noise, that should be proportional to the effects
associated with the flow of a single droplet, should decrease in
relation to the overall dissipation in a large system. From this, and
from the fact that the number $\Omega$ of possible trajectories explodes
in a factorial fashion with $N$, one can expect that in a large system the
chance of observing any particular pattern, as well as the chance of coming
back to a pattern once abandoned (due to noises or disturbances) should
be vanishingly small.

In order to verify these predictions, we performed an experiment
using the same technology as in the recent experiment that confirmed the
ideal model\cite{olgierdLOC}; the only difference was that the two
channels forming the loop were much longer (about 180 mm -- more
than 400 times longer than they were wide). These channels could
comprise hundreds of droplets at a time. Quite surprisingly, our
predictions proved completely wrong: instead of observing an erratic
creeping of the system over the whole space of allowed configurations,
the system always settled into the ``maximally
queued'' pattern, as in \mbox{Fig.~\ref{fig:im1} (d) -- Fig.~\ref{fig:im1}(e)}.
This behavior was neither dependent on the frequency of feeding the
droplets into the loop nor on the droplet size: as long as droplets
were able to pass the junction without breaking-up, the long queues
always appeared. We also observed that this behavior did not depend on
the method of generation of droplets: we reproduced the same results
with i) a well-tuned Droplet On Demand (DOD) system\cite{Churski2013}
that produced a sequence of equally spaced monodisperse droplets, ii)
with an improperly vented DOD valves producing droplets with a
considerable variation in their size, and even iii) without an active
control of the generation of drops, when both their volume and spacing
between the drops changed in response to the changing load of the
system. Although initially (if starting from the empty channels) the
droplets tended to form patterns similar to these
from \mbox{Fig.~\ref{fig:im1} (b) -- Fig.~\ref{fig:im1}(c)}
(i.e. ``chopped''), the long trains of drops always appeared within
minutes and persisted, no matter how long the experiment ran.

Puzzled by this result we considered the following questions: (i) can
the observed dynamics be explained within the ideal model? If not, (ii)
what kind of nonlinearities or other corrections should be taken into
account to find the possible reason in a minimalistic yet plausible
way? If yes, (iii) why there was not a single report of this problem in
the literature related to a wide range of simulated microfluidic
networks?

The last question proved most helpful in searching the literature for
aspects that have not been considered. Most reports on modeling flow of
drops in microfluidic networks focus on systems with channels of varied
lengths yet always of constant cross-section. This seems a natural
choice, one for the clarity of analysis, second for that it is typical
and experimentally easiest to prepare chips of (nominally) constant
height and width of the channels. If the cross-section of the channels
were truly constant, it would be necessary to look for the reason of
queuing in subtle and little known effects, such as e.g.
distance-dependent hydrodynamic interactions between
droplets\cite{SchindlerResistance}. From auxiliary measurements we found that
this effect was negligible under the conditions of the experiment
described here. Moreover, we observed the queuing dynamics regardless
of inter-droplet separation even though the
cooperative effects, if any, should be negligible for large intervals between
droplets.
We thus focused further analysis on the effects of the
varied cross-section on the dynamics of flow of drops through networks.
An extra reason to follow this path was the fact that both the process
of milling long channels and bonding chips in a hot press are prone to
systematic deviations of the depth of channels -- due to thermal expansion
of spindle in milling and due to non uniform temperature and
pressure field in a press.

\section{Theory and simulations}
In the case of flow of simple, Newtonian fluid, the variation of
cross-section of the channel along its length cannot produce any
variation in time. Even though the local pressure gradient depends on
the local geometry, any given spot along the channel is always filled
with the same liquid. Yet, when the fluid is complex and
non-homogeneous, such as e.g. a suspension of droplets, the local
pressure gradients will depend on the content of the channel at any
given point. In consequence also the total hydrodynamic resistance of
the channel may depend on the position of droplets along its length,
and change in time.  Even for a simple liquid the hydrodynamic
resistance ($R$) is very sensitive to changes in the transverse dimensions
(say $d$) of the channel. According to the Hagen-Poiseuille law, reduction
of the diameter of the duct from $d$ to  $\gamma d$ ($\gamma<1$)
changes the resistance by a factor $\gamma^{-4}$.
In the case of an immiscible droplet, this scaling is much
more complicated: not only the droplets' cross-section
area decreases by a factor $\gamma^2$, but also its length increases
by a factor $\gamma^{-2}$.
Further, its linear speed of flow must -- by conservation of mass
-- increase approximately $\gamma^{-2}$ times.
In a very rough approximation, the
elongated droplet of high viscosity may be described as a slug of
Hagen-Poiseuille flow of the length $\Delta l$, with effective
viscosity $\eta_\text{eff}$ being the difference between the viscosity
of the discrete and the continuous liquids,
$\eta_\text{eff}=\eta_\text d-\eta_\text c$.
Within this crude approximation, the resistance introduced into the channel
by a droplet is proportional to $\Delta l/d^4$, and the change
associated with reduction of the diameter of the pipe scales as $\gamma^{-6}$.
For example, a mere 10\% decrease of the diameter of the pipe results in a
23\% increase in the length of the drop and in almost doubling of the
resistance the droplet adds to the resistance of the channel. The above
estimation is rough -- the only important insights that we draw from
it are that i) the total resistance to flow in a microfluidic channel
increases when droplets move into a narrower segment, and that ii) this
effect may be significant.
Instead of dealing with unknown dependence of these additional resistances
on cross-section of channels, flow rates and size of droplets,
we simply assume, that the resistance of a droplet is given by $r_\text{up}$
in the upstream segments, and $r_\text{down}$ in the downstream segment (see
Fig.~\ref{fig:im3}). As the resistances of individual droplets may differ
(due to emulated noise in simulations or polydispersity of droplets in
experiments), we will rather use the ratio of these resistances:
\begin{equation}
\label{eq:alphageneral}
\alpha = \frac{r_\text{down}}{r_\text{up}}\;,
\end{equation}
having in mind that $\alpha$ may be equal to $r_\text{narrow}/r_\text{wide}>1$
in loops narrowing downstream, or $r_\text{wide}/r_\text{narrow}<1$ in
the opposite case.

\begin{figure*}
\centering
\includegraphics[width=0.92\textwidth]{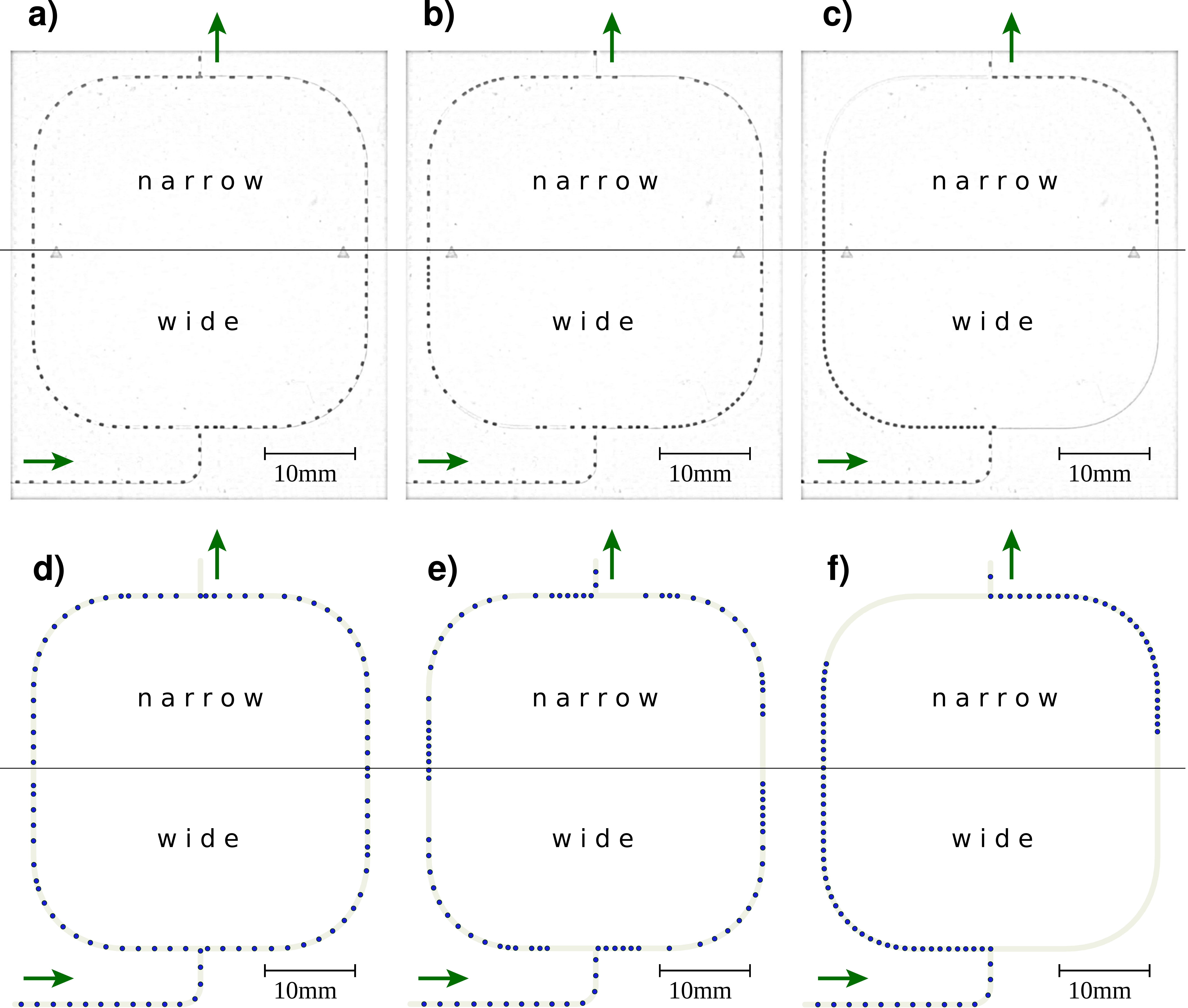}
\caption{\label{fig:comp} (color online)
Top panel (a--c): photographs of the experiment with evolving pattern of
droplets in the loop with narrowing channels: (a) -- at the beginning,
(b) -- intermediate, (c) -- stationary state.
Bottom panel (d--f): snapshots of corresponding simulations.
}
\end{figure*}

Networks comprising segments of slightly different cross-sections may be
simulated without adding new features to the
Schindler's model; it is enough to divide channels into
subsections, each having its own geometry and its own resistance
multiplier for droplets, and simulate the divided channels as being
connected in series. The notion of  slightly different cross-sections
requires clarification: we use this term to emphasize the fact (or
assumption) that the dynamics of the system is dominated by collective
effects of additional resistance carried by droplets and not by local
effects of passing through the junctions between channels of different
cross-section. Presence (and motion) of droplets in
such a section involves additional pressure drop due to capillary
effects\cite{JensenClogging}. This transient pressure drop could be
taken into account for reliable analysis and simulation, but for the
simplicity -- since we only want to demonstrate that the queuing
phenomenon may be related to a non constant cross-section of channels
-- we will neglect it.

Therefore, using the simple model and neglecting capillary effects on
channel contractions, we tested several simple modifications of a
classic asymmetric loop. We introduced narrowed sections of channels
equal to either half or the whole length of the single parallel section
and we varied their location. Bottom panel of Fig.~\ref{fig:im3} shows
the configurations that we tested.
The segments marked with the thin line are narrower than those drawn
with the wide solid line. 

We observed that these systems varied strongly in their tendency to form
long queues: some of the geometries exhibited queuing (...LLLLRRRR...),
some seemed to suppress queues (i.e. to prefer frequent switching between
L and R), the behavior of others depended strongly on the initial conditions.
Specifically, geometries B and C where neutral in the sense,
that they exhibited the full range of combinatorial patterns,
as in the unmodified case A.
Geometries D, K and M tended to suppress queuing -- even if the
simulation had been started from a ``perfect queue''. The cases F
and G were difficult to classify; for some specific sets of starting
configurations the preferred patterns were
``queue-like'' yet the length of queues were
far from maximum. Finally, L, N, and most of all E, formed queues of
the maximum possible length (i.e. longer than half of the length of
the channels).

\begin{figure*}
\centering
\includegraphics[width=1\textwidth]{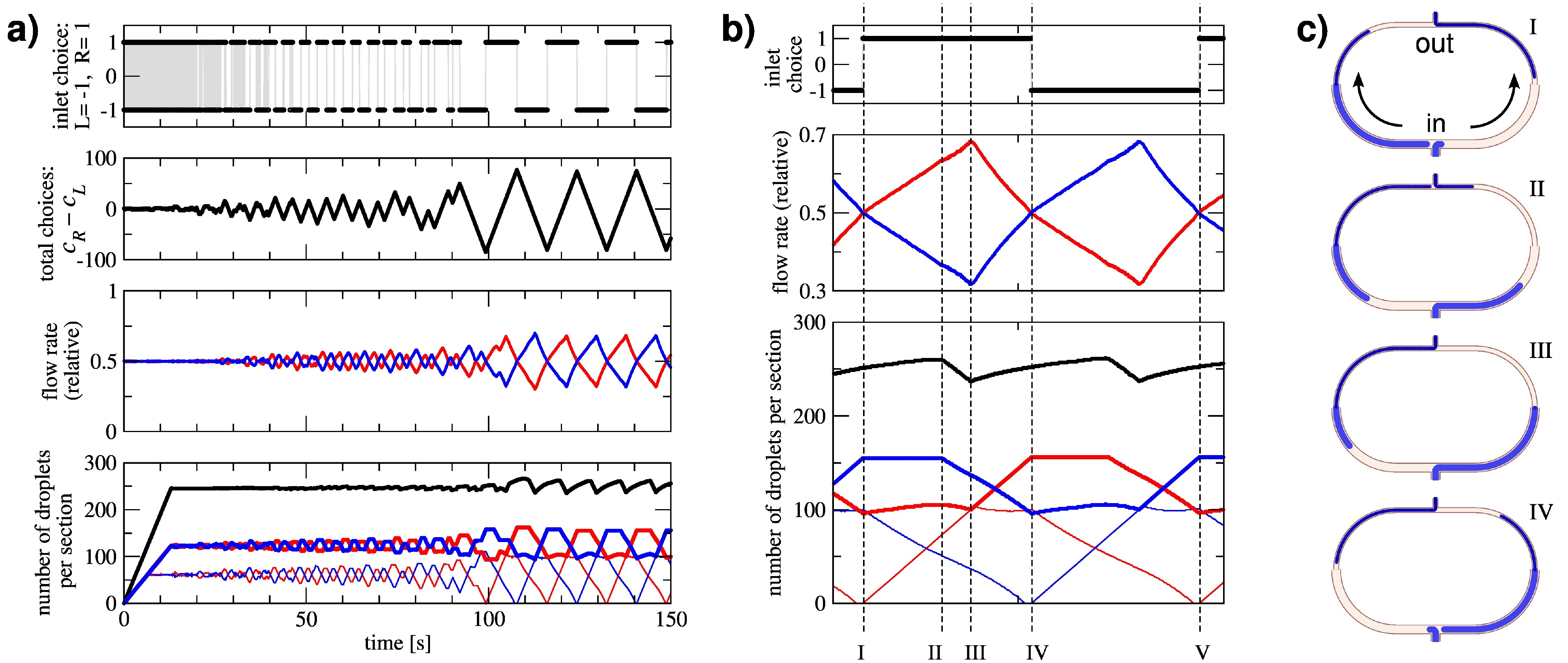}
\caption{\label{fig:im4} (color online)
Onset of oscillations in the symmetric loop
comprising sections of narrowed channels from a computer simulation.
(a) Time dependence of key variables in the simulation
started from an empty system. The ``inlet
choice'' signal (top panel) is assigned the value $+1$ if
a droplet flows to the right channel and $-1$ if to the left.
Second panel plots the difference $C(t)=C_\text{R}(t)-C_\text{L}(t)$
between the number $C_\text{R}(t)$ of droplets that had chosen
the right channel and $C_\text{L}(t)$ for the left one.
Third inset shows the flow rates through the arms
(solid blue for the left channel, dashed red for
the right one) normalized by the total volumetric flow through the
loop. Fourth chart plots the momentary numbers of droplets in the whole
loop (thick black solid line), in the right arm (thick solid red line)
and in the left arm (thick dashed blue line). The thin lines show the
number of drops in the initial (not narrowed) sections of these
channels. (b) Magnified part of the same plots, comprising
single, fully established cycle of oscillation, with characteristic
instants of time marked by roman numerals. (c) Positions of
droplet queues at these instants: (I) Due to a long series of droplets
previously loaded into the left channel, its resistance becomes higher
than of the right channel; new droplets start to enter the right one
(II) front of the queue in the left channel reaches the end of the
loop; (III) front of the queue in the right channel approaches the
narrowed (and momentarily empty) section; the flow rate in the right
channel achieves maximum value, while rate of flow in the left channel
is minimum; (IV) symmetric to (I); (V) completion of the cycle -- the
system comes back to the configuration (I).
}
\end{figure*}

\subsection{The onset of oscillations}
In the following analysis we focus on systems with narrowed sections
positioned at the downstream half of the parallel channels
(as in Fig.~\ref{fig:im3} (E)). For simplicity we consider a symmetric loop
with both parallel sections of equal length and having identical segments of
reduced lumen.
Examples of the process described here are shown in Fig.~\ref{fig:comp} --
the three snapshots of experiment (details of the experiment are
provided in Sect. \ref{sect:exp})
and corresponding computer simulation illustrate the evolution of patterns
at the beginning (just after first droplets reach the end of the loop that
was initially filled solely by the continuous liquid), in an intermediate
state (as the queues start grow), and with fully developed queues.

When the first droplet enters the (initially empty) loop, it may flow into
any of the arms. In real systems the choice is predetermined by inevitable
deviations from the perfect symmetry of channels and T-junctions, and in
simulations the decision must result from the details of algorithm and
from the finite precision arithmetics.
This droplet increases the hydraulic resistance of the chosen channel
by $r_\text{wide}$.
The subsequent droplet will then choose the opposite branch, and once it
flows in, it increases the resistance of the second channel by the same
amount, balancing the rates of flow through each of the arms. If the
symmetric loop did not comprise the narrowed sections, this simple
mechanism would lead to a ``chopped''
state with alternating L/R trajectories of the droplets. This
uniform configuration of droplets
can be sustained \textit{ad infinitum}: as a drop leaves a branch and
decreases its resistance, the next droplet will flow to the same arm of
the loop. Interestingly, the introduction of a narrowing at the
downstream termini of the arms destroys the stability of the
``chopped'' state. When a droplet
passes into the narrowed section, the drops' contribution to the
resistance of the channel increases
from $r_\text{wide}$ to $r_\text{narrow}$, so that the excess of
flowing resistance in this branch is amplified by a factor
$\alpha = r_\text{narrow}/r_\text{wide}$ (see Eq. \ref{eq:alphageneral}).
This increase must be compensated by new droplets entering the opposite
branch and since they introduce smaller resistive contributions, one
droplet does not balance the inequality of resistances of the two
branches. The combination of the change in the resistive charge of the droplets
and the delay between a decision taken at the inlet and its amplification
at the narrowing, provides for the instability of
the chopped states: the perturbation will be larger and larger with
each cycle, finally leading to the oscillation of the maximum possible
amplitude.

Fig.~\ref{fig:im4}(a) shows an example of development
of queues in a simulation with $\alpha=2.44$. In Fig.~\ref{fig:im4}(b)
we graphed a zoom on the dynamics of the system in the fully
developed queued state. Top panels of these plots show a discrete
function taking on two distinct values: $+1$ if a droplet flows
to the right arm (R), and $-1$ if to the left (L).

Integrating this signal over time leads to the function $C(t)$
plotted in the second panel of Fig.~\ref{fig:im4}(a).
This function may be interpreted as the difference between the
numbers $C_\text{R}(t)$ and $C_\text{L}(t)$ of droplets that flew
to the right and left channels respectively (from the beginning of
the simulation). Due to the symmetry of the loop, both $C_\text{L}(t)$
and $C_\text{R}(t)$ increase with the same average speed, and
$C(t)=C_\text{R}(t)-C_\text{L}(t)$ oscillates around zero.
However, the amplitude of changes of $C(t)$ provides a measure of queuing.
In a perfectly chopped state $C(t)\in(-1,1)$ because droplets enter
the channels in alternation. In a queued state $C(t)$ departs far
from zero because long trains of droplets follow the same trajectory.
We normalized the rates of flow through each of the branches by the
total volumetric flow rate -- hence the rates of flow through each
of the two branches oscillate around the value of 0.5 (third panel
of Fig.~\ref{fig:im4}(a) and second panel of Fig.~\ref{fig:im4}(b)).
At bottom panels we also plot the numbers of droplets residing in
the right/left arm of the loop at any given instant of time,
and the corresponding numbers of droplets in the initial
(not narrowed) sections.

Since the simulation starts with empty channels, at the beginning
droplets enter both channels in an alternating manner; both trains of
droplets move downstream their arms at almost the same speed and reach
the narrow segments almost in the same time; then, similarly, they
reach the end of the loop. In the absence of any noise, this behavior
could continue infinitely; locking the system in this metastable state
is possible because the loop is symmetric. The onset of queuing
requires a small perturbation or a finite level of noise. We will
discuss the role of noise in detail in the next section. Interestingly,
the growth of queues is not monotonic. Usually two or more shorter
chains grow independently to finally combine into the queue of the
terminal (maximum) length.

\begin{figure*}
\centering
\includegraphics[width=1\textwidth]{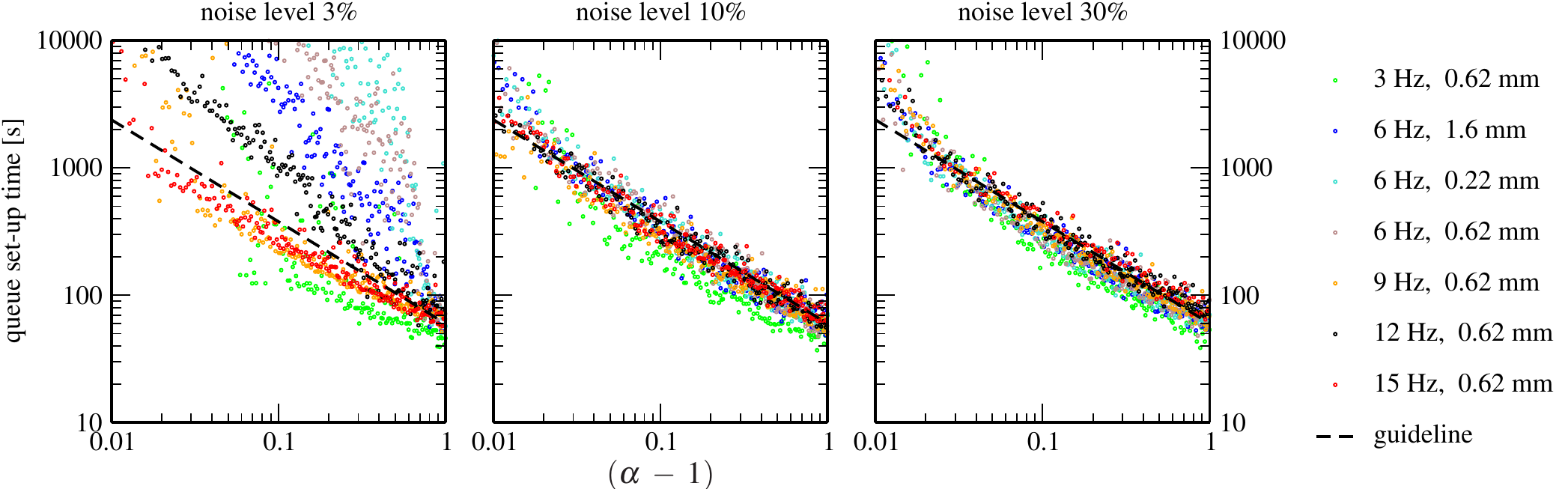}
\caption{\label{fig:im5} (color online)
Time of approach to a fully developed oscillation, as a
function of the factor of increase of the hydrodynamic resistance of
droplets entering the narrowed section of the channel. The three graphs
differ in the noise level: from left to right the amplitude of noise is
3\%, 10\% and 30\% of the mean interval between drops and of the mean
resistance of the drops. Simulations were performed for varied
frequency [Hz] of feeding droplets into the loop and for varied
resistance equivalent length [mm] of the droplets (see the legend). The
dashed line is a power function proportional
to $(\alpha-1)^\beta$, with $\beta=0.8$. All simulations
were performed for a symmetric loop with arms of length 20 mm, divided
into two equal segments of different cross-section. The linear velocity
of feeding droplets (in the common inlet channel) was set to 3.1 mm/s.
The resistance equivalent lengths are given with respect to the initial
(wide) segments of the loop.
}
\end{figure*}

At the first sight the fully established queued state in the system
comprising the narrowed sections (Fig.~\ref{fig:comp}(c) and
Fig.~\ref{fig:comp}(f), Fig.~\ref{fig:im4}(b)--Fig.~\ref{fig:im4}(c)),
resembles the queued pattern from
\mbox{Fig.~\ref{fig:im1}(d)--Fig.~\ref{fig:im1}(e)}
in a symmetric loop without any constrictions.
There are, however, important differences. First, in the system
without constrictions, the queued pattern is solely
\textit{one of multiple} degenerated stationary states
and it can thus be easily destroyed by a random perturbation.
In the system with the narrowed sections the queued state is the
\textit{only} stable stationary state and is robust against noise.
Second, in the constant cross-section system both the number of droplets
and the rates of flow are -- to within a single droplet and its effect of
the flow rate -- constant in each branch. In the narrowing loop both of
these quantities change significantly within each cycle. The magnitude
of these changes increases with the degree of modification: the
narrower is the cross-section of the modified segments, the larger are
these deviations. Moreover, exactly these cyclic variations contribute
to the stability of the queued state: the large momentary differences of
the flow rates determine the trajectories of drops entering the loop.
Random perturbations may change the trajectories of the droplets only
at the ends of the queues -- as only at the moment of switching the flow
rates in both arms are of similar value. Third, in the unmodified
system, the droplets formed mutually complementary patterns in the two
arms: the positions of droplets in one channel matched positions of the
gaps in the second (see the thin lines in Fig.~\ref{fig:im1}). In the
modified system this rule no longer holds -- the lengths of the queues
are substantially longer than half the length of each channel
(see Fig.~\ref{fig:comp}(c) and Fig.~\ref{fig:comp}(f), Fig.~\ref{fig:im4}(c))
and are not complementary. For example in configuration II in
Fig.~\ref{fig:im4}(c) the queues from the two arms flow into the outlet
at the same time, possibly leading to mutual collisions of droplets leaving
each of the arms. This lack of complementarity causes both the total number
of droplets in the loop and the pressure drop at constant flow rate conditions
(or the flow rate in constant pressure) to vary within each cycle.

\subsection{Suppressing oscillations}
\label{sect:supp}
If, on the other hand, narrowed sections are positioned at the
upstream half of the parallel channels (as in Fig.~\ref{fig:im3}(D), (K)
and (M)), the oscillations not only do not occur, but also will be suppressed
(if the queued state is artificially created by the initial configuration).
It means that the same loop, traveled by the same sequence of droplets
(i.e. with the same flow rate, volumes, and separation between droplets)
will behave completely differently after switching the direction of flow.
In particular, reversing the direction of flow in the system shown in
Fig.~\ref{fig:comp}, will lead to patterns with low number of the same choices
(i.e. LL or RR) in order.
Mathematically, the reversal of the direction of flow means changing
$\alpha$ (see Eq.~\ref{eq:alphageneral}) from $r_\text{narrow}/r_\text{wide}$
to $r_\text{wide}/r_\text{narrow}$.
If $\alpha>1$ corresponds to amplifying fluctuations of ``flowing resistance''
and growing oscillations to highest available amplitude, reversal of the flow
means that the oscillations become damped. Nevertheless, it does not simply
mean the reversal of the direction of evolution nor the substitution of the
target of the evolution from perfect queuing (...LLLLRRRR...)
to perfect chopping (...LRLRLRLR...).
For $\alpha>1$ the evolution leads to uniquely defined state of the perfect
queues (perhaps with some minor deviations only at the very ends of the queues,
caused by high levels of noise).
In contrast, reversing the flow ($\alpha<1$) does not lead to a single, unique
state, but to a family of microstates that distribute droplets between two arms
more or less homogeneously.
These microstates can easily mutate into each other
bacause (in contrary to the queues for $\alpha>1$) the system is always almost
balanced -- the difference of flow rates between arms is as small as the effect
of single droplets, so that a random permutation may be located in any spot
along the sequence.
Thus in the case of $\alpha<1$ the evolution seems to escape
from queue rather than run towards any particular, highly chopped pattern.

Although the confirmation of this finding in computer simulation is
straightforward, as the simulation may be started from any initial
configuration of droplets, experimental tests require more caution.
Typical experiment starts from empty loop and then continue to more or less
uniform distribution of droplets. Therefore, in order to prove that queuing
is suppressed, the queue must be created artificially.
We did it by freezing the flow in one of the arms (by placing a piece of dry
ice above the chip), so that droplets could not enter the blocked channel.
After removing the ice and melting the frozen oil, the channel was again
open for flow, but the artificially created queue was not stable and evolved
to a chopped state after several cycles (see the details in
Sect. \ref{sect:exp}, and in particular in Fig.~\ref{fig:im7}(b)).

\subsection{Effects of noise and of geometry on formation of queues}
As discussed above, the onset of queuing usually requires a random
perturbation that may change the
``ideal'' pattern of L and R choices.
In a small experimental systems it is likely that the experimental
noise is too small to induce a change of the stationary pattern of
trajectories. In a large system, with a large number of droplets, their
time of residence in the parallel arms is long enough for small
perturbations to accumulate. In simulations, we added artificial noise
by randomizing the intervals between subsequent droplets entering the
loop and by randomizing the volume and charge of resistance of the
droplets. Intervals between the droplets and their resistances were
drawn from a symmetric triangular distribution centered at the required
mean value with ranges given as a specified percentage of the mean. For
example, in the simulations plotted in
Fig.~\ref{fig:comp}(d)--Fig.~\ref{fig:comp}(f) and Fig.~\ref{fig:im4},
the emulated noise was $\pm 10\%$ of droplet resistance and $\pm 10\%$
of inter-droplet interval. The standard deviations of the corresponding
distributions were $\sqrt{6}$ times smaller, i.e. $4.1\%$ of a mean value.

We used simulations to test the minimum amplitude of noise needed for
initiating the process of queuing. In order to quantify how quickly the
queues grow, we constructed a quantity reflecting the extent to which
the state of the system resembled a ``perfect
queue'' comprising a maximum number of droplets in a row.
An exact estimation of this maximum length is not straightforward, but
it is of the order of the maximum number of droplets in the whole loop.
This number is also difficult to calculate exactly yet using the
``mean field approximation'' for an
unmodified loop, we obtain the following estimation:

\begin{gather}
\bar N_\L=\frac{L_\L \bar f_\L}{\bar v_\L}\quad ,\quad\quad
\bar N_\R=\frac{L_\R \bar f_\R}{\bar v_\R}
\end{gather}
where $\bar N_i$ is the average number of droplets, $L_i$ is
the length, $\bar f_i$ is the average frequency of
entering droplets, and $\bar v_i$ is the average velocity
in the $i$-th channel ($i = \L$ or $\R$). For a symmetric, unmodified
loop ($L_\L=L_\R=L$) the corresponding values are equal:
\begin{gather}
{\bar v_\L} = {\bar v_\R} = \frac{Q}{2 A}\quad ,\quad\quad
{\bar f_\L} = {\bar f_\R} = \frac{f}{2}\\
\label{eq:cyclength}
\bar N_\tot = \bar N_\L + \bar N_\R = \frac{2 A L f}{Q}
\end{gather}
where $Q$ is the total volumetric flow rate, $f$ is the frequency
of feeding droplets into the loop, and $A$ is the cross section area of
the channels. Here we assume that $A=w\cdot h$ is constant along channels
while the increase of the hydraulic resistance introduced
by the droplets results from modification of the aspect ratio $w/h$
of the width ($w$) and height ($h$) of the channel. This way the resistance
of droplets is changed, but their length, separation and velocity are not.

The ``perfect queue'' should
contain at least $\bar N_\tot/2$ droplets flowing one after another
into the right channel and the same number of drops then flowing
into the left arm. We monitor the interval required for the onset
of the oscillations as the interval from the start of simulation
until the first occurrence of $\bar N_\tot/2$ subsequent droplets
following through the same channel.

Fig.~\ref{fig:im5} shows the time required for the onset of oscillations
as the function of $\alpha$ (defined in Eq. \ref{eq:alphageneral}).
We run the simulations keeping constant flow rate and varying
the frequency of feeding droplets into the loop, the volume of
the droplets, and the amplitude of noise (same for the intervals
between the drops and the resistance they carried). We note that the
time required for the onset of oscillations depends very weakly on the
resistance of droplets and on the frequency of drops entering the loop.
It is possible, that this dependence, visible especially at low noise
level on the left graph in Fig.~\ref{fig:im5}, results mainly from the
hidden influence of these quantities on the overall fluctuations in the
flux of hydrodynamic resistance through the system. This is because
we define the amplitude of the noise in relation to the resistance of
droplets or frequency of feeding the droplets; increasing these
parameters increases also the amplitude of fluctuations. These effects
play a crucial role when the level of noise is small, and $\alpha$ is too
small to induce permutation of a pattern without a random perturbation.

\begin{figure}
\centering
\includegraphics[width=0.46\textwidth]{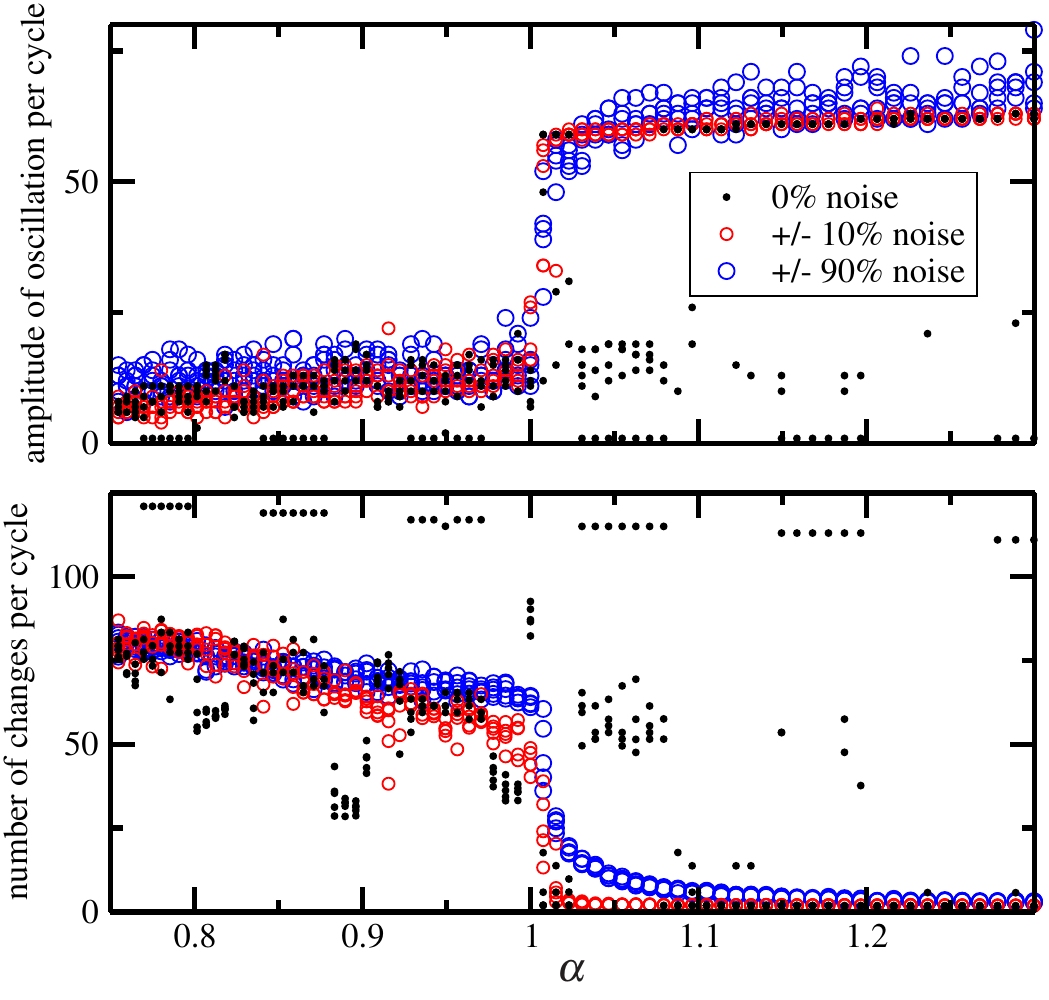}
\caption{\label{fig:im6} (color online)
The amplitude of oscillations and the number
of changes of the trajectories of droplets within one cycle as a
function of $\alpha$. Data shown for long, randomly initialized simulations
of the symmetric loop containing about 120 droplets. The amplitude of
oscillations assumes high values in states with long queues, while the
number of changes of the trajectories assumes small values in the
queued state. From the graphs it is clear that when the downstream
halves of the channels induce higher hydrodynamic resistance of
droplets ($\alpha>1$) queues form, while the reduction of the resistance of
introduced by the droplets in the downstream sections ($\alpha<1$) suppress
queues. This behavior is stabilized by a moderate level of noise.
}
\end{figure}

\subsection{Effects of noise and of geometry on the quality of queues}
Apart from the onset of oscillations, also the
``quality'' of queues is of interest.
In order to test the stability of the queued states we monitored to
what extent the state of the system resembled the perfect queue. We
note that the sole number of identical subsequent decisions overlooks
states that differ only very slightly from a ``perfect
queue''. For example for $\bar N_\tot/2=12$ the
string LLLLLLLRLLLLLLLL will be rejected despite of more
than 12 L choices almost in a row. In order to take into account
these imperfections we introduced the following criteria:

(i) the amplitude of oscillation, defined as the difference
between maximum and minimum of value of $C(t)$ within a cycle.
To avoid ambiguities in identification of the cycle and
calculating its length, we use its rough estimate $\bar N_\tot$.

(ii) The number of changes from L to R or vice versa during the cycle.
Perfect queuing should yield 1 or 2 changes depending on whether the
actual length of the cycle is longer or shorter than the
estimate $\bar N_\tot$.
In contrast, perfectly chopped state (in a symmetric loop) produces
maximum possible value, which is equal to the length of the cycle.

\begin{figure*}
\centering
\includegraphics[width=0.99\textwidth]{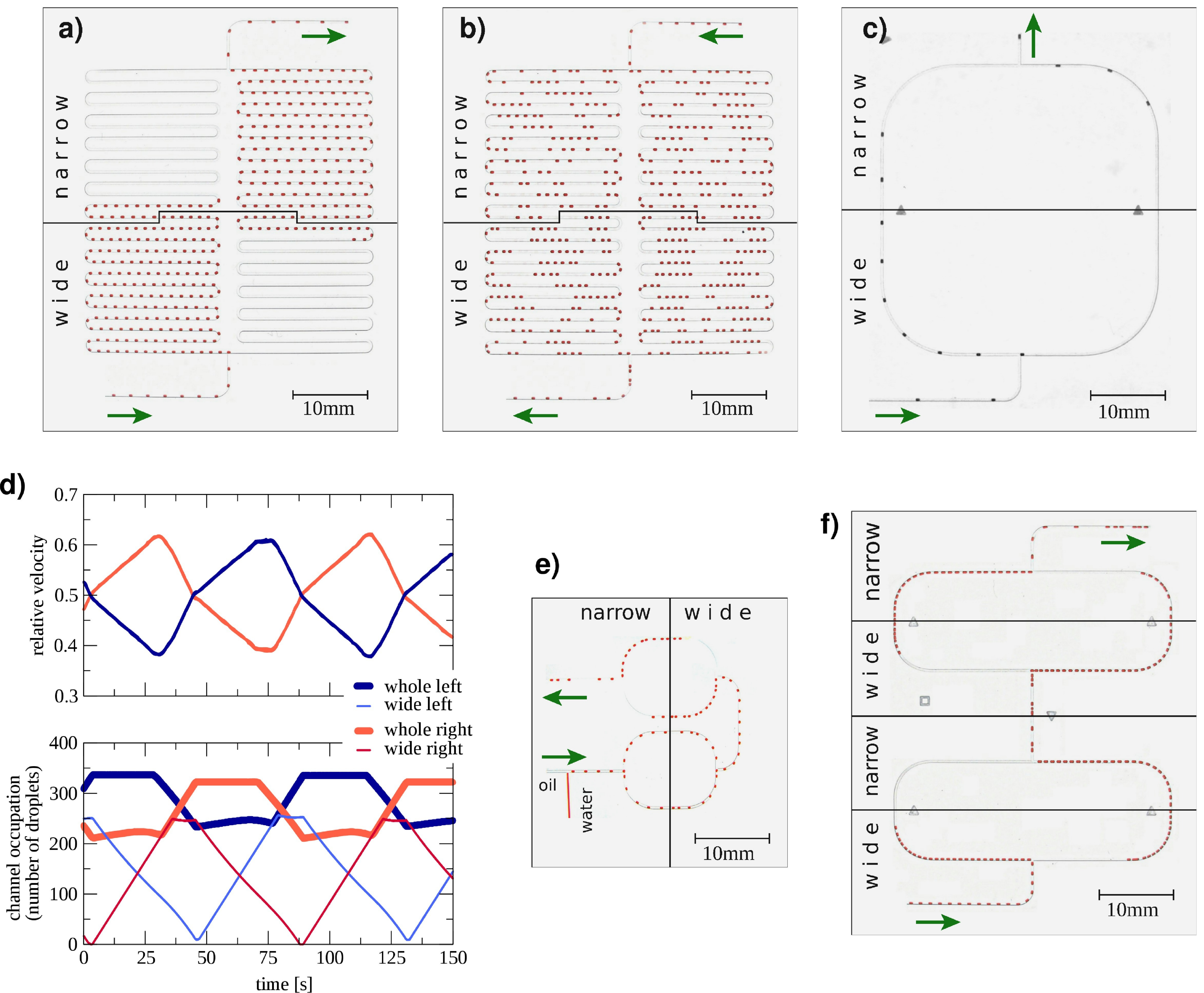}
\caption{\label{fig:im7} (color online)
Snapshots of the experimental systems after
long time of evolution at constant conditions. Direction of flow is
marked with arrows. Black lines demarcate segments of channels with
different cross-section, labeled as ``wide'' (width $\times$ height =
370 $\times$ 400 ${\mu}$m) and ``narrow'' (370 $\times$ 340 ${\mu}$m).
(a) Queuing state of the long, symmetric loop. The ``wide'' channels
(in fact ``deep'') are 230 mm long, and the ``narrow'' (shallow)
are 250 mm long.
(b) The same chip with reversed flow direction.
(c) Queuing state in the symmetric loop of moderate length (66
mm per channel, equally divided between shallowed and deepened
sections). This is the same chip as in Fig.~\ref{fig:comp}(a)--
Fig.~\ref{fig:comp}(c), but there with larger separation between droplets.
(d) Dynamics of queuing state in the chip a).
(e) Two identical, short loops in series; the first one is
flown from shallow to deep, the second the opposite. Each arm is 18 mm
long. (f) Two identical loops in series, both flown from ``wide'' to
``narrow'', each arm is 55 mm long.
}
\end{figure*}

Fig.~\ref{fig:im6} shows how these two measures of queuing depend on the
factor $\alpha$, defined in Eq. \ref{eq:alphageneral}. Each point in these
graphs represents a data point from a simulation, started from random initial
conditions and recorded after a long time (10 000 droplets). In these
simulations $\bar N_\tot=116$. We varied the level of noise (0, 10 or 90\%)
and $\alpha$, all the other parameters held constant.
$\alpha<1$ corresponds to loops comprising downstream halves of the two
parallel channels wider (less resistant) than the upper ones, and $\alpha>1$
codes for systems with narrowed sections. At moderate level of noise
(10\%, or 4.1\% in terms of the standard deviation) both criteria indicate
perfect queuing already for values of $\alpha$ as close to 1 as just
$\alpha>1.02$. The slight increase of the amplitude of oscillations is due to
the hidden dependence of both $N_\tot$ and the actual length of the queues on
the cross-section area of modified sections (changing together with $\alpha$).
For the large level of noise (90\%) and $\alpha>1$, there are visible
imperfections in the queues, increasing for $\alpha\approx 1$ .
In a stark contrast, when the downstream sections provide for a lesser
resistance introduced by the droplets then the upstream sections of the
channels (i.e. for $\alpha<1$), queuing is suppressed, regardless of the
level of noise. Interestingly, at a high level of noise the number of
changes per cycle is close to half its length, just as if the sequence
of L/R choices was random. On the other hand, the amplitude of oscillation
measured for real random binary strings (resulted from Bernoulli process,
as repeated coin flipping), is higher than that recorded in the simulation.
We attribute this fact to the mechanism of suppression of long queues
(see Sect. \ref{sect:supp}).

It is also very interesting to note, that in the absence of noise (0\%)
all the above regularities become less evident. For example, for $\alpha>1$
queuing may be overcome by regular memorized
patterns\cite{olgierdLOC} (notice the characteristic structure of
bands in the dependence on $\alpha$). Locking the system in a
``perfectly chopped'' state is still
possible, although it requires very special initial conditions, as
starting from the empty loop. Interestingly, suppression of long queues
in widening channels ($\alpha<1$) is still efficient in the absence of noise:
even if simulation starts from perfect queues, they disappear. This is
because the imbalance needed for changing a pattern comes collectively
from all droplets, if starting from a queue, and from one droplet only
(plus fluctuations / noise), if starting from the chopped pattern (as
well as from the empty loop). 

\section{Experimental results}
\label{sect:exp}
Our predictions have been fully confirmed in experiments. We
manufactured four different microfluidic chips, each comprising two
kinds of channels: ``narrow'' and
``wide''. Despite these names, used
for the consistency, the channels, milled in a slab of polycarbonate
and sealed thermally\cite{DominikaBonding}, are of the same width of 370
${\mu}$m and differ only in height: 340 ${\mu}$m for the
``narrow'' channels and 400 ${\mu}$m
for the ``wide'', with smooth, 1 mm
long, transitions between them. As working liquids we used hexadecane
with 0.5\% (w/w) of Span80 surfactant and water dyed by 8\% (w/w) of
red ink (Encre Rouge / Waterman, Paris). For these liquids and channel
geometries we estimate the corresponding value of $\alpha$ for droplets of the
volume of 60 nl to be $\alpha\approx 2$ -- for the linear velocity of droplets
close to 5 mm/s, as used in the presented experiments. The smooth (1 mm long)
transitions between the wide and narrow segments of the channels
produce very small changes in the curvature of the droplet and the
capillary pressure (of the order of tens Pa). The chips comprised one
or two symmetric loops and one or two T-junctions for generating
droplets. Using two generators placed on opposite ends of the loop
enabled us to invert the flow direction so that droplets could enter
the loop starting from shallowed segments as well as from the deepened
ones. Droplets could be created passively by introducing constant rates
of flow of the two liquids, yet in order to increase the range of
frequencies and sizes of droplets we used DOD (droplet on demand)
technique based on pressurized containers equipped with valves and long
capillaries between the valves and the chip\cite{Churski2013}.

The chip presented in Fig.~\ref{fig:im7}(a)--Fig.~\ref{fig:im7}(b)
(microphotographs) and in the graph in Fig.~\ref{fig:im7}(d), comprises
a long loop with arms 48 cm long, each including 25 cm of deepened
segment and 23 cm shallowed. As long as droplet separation was large enough
to avoid mutual collisions at the inlet bifurcation\cite{CollisionPanizza},
this chip always behaved in agreement with our predictions: droplets flowing
towards shallower segments formed perfect queues, as in Fig.~\ref{fig:im7}(a),
whereas in the opposite direction the patterns were quasi-random, avoiding
long queues -- typical example is shown in Fig.~\ref{fig:im7}(b). Regardless
of the initial state, the time of formation of the perfect queues was not
longer than 10 cycles.

Fig.~\ref{fig:im7}(d) shows cyclic changes in the flow rates and in occupation
of channels in the system from Fig.~\ref{fig:im7}(a), with a fully developed
queue. These experimental data can be qualitatively compared with the plots
in Fig.~\ref{fig:im4}.

Fig.~\ref{fig:im7}(c) presents queuing in a symmetric loop (the same as in
Fig.~\ref{fig:comp}(a)--Fig.~\ref{fig:comp}(c)) with arms of length of 66 mm
(33 mm shallowed and 33 deepened).
In Fig.~\ref{fig:comp}(a)--Fig.~\ref{fig:comp}(c) the center--to--center
separation of droplets at the inlet was set to 1.47 mm -- as closely as
possible to the limit of the collision regime\cite{CollisionPanizza},
and in Fig.~\ref{fig:im7}(c) it is 10 mm.
Supporting video\footnote{See the supplementary material} shows the evolution
of this system from the empty loop up to the fully developed queues.

Fig.~\ref{fig:im7}(e)--Fig.~\ref{fig:im7}(f) exemplify the use of the
understanding of what geometries promote / suppress queuing in constructing
microfluidic systems. Each of the systems comprises two identical loops
connected in series. In Fig.~\ref{fig:im7}(f) both loops are traversed
from the deepened to the shallowed segments, so that the queues are visible
in both the loops. In Fig.~\ref{fig:im7}(e) only the second loop is oriented
for queuing; the first loop is oriented so that the flow proceeds from the
shallowed segments towards the deeper ones, so that queuing is suppressed.
Unlike the long and moderate loops from
Fig.~\ref{fig:im7}(a)--Fig.~\ref{fig:im7}(c), these from Fig.~\ref{fig:im7}(f)
could present also patterns other than perfect queuing; it is clear
that noise is required to induce transitions in loops of such a small size.
Notice that the effective margin of noise in the second loop is elevated
by varying intervals between droplets at the output of the first loop.

\section{Conclusions}
We believe that our work may shed new light on the dynamics of the
flow of droplets through microfluidic networks. We demonstrated that a
theoretical description\cite{olgierdLOC,Glawdel,PanizzaComplexPRL,
JeanneretHamiltonian,AgentBased,Iranczycy}
built on the ideal model\cite{schindler:08} does not describe the effects
caused by varying cross-section of channel and the noise that
is always present in real experiments.
Adding a small \emph{intentional} modification of channels may be used
for turning a chaotic system into a device that reproducibly either
distributes droplets uniformly between the parallel channels, or forms
long queues of droplets. The experiments confirmed our numerical
predictions. We found that droplets flowing through a microfluidic loop
comprising two channels that narrow towards the outlet, group
themselves spontaneously into queues of maximum possible length. This
behavior is independent on any parameters of flow, provided the
droplets neither break at the bifurcation nor collide there. Although
the mechanism of formation of queues is related to noise, once
established, the queues cannot be destroyed even by large amplitude of
random perturbations in the intervals or volumes of droplets. We also
demonstrated in simulation and experiments that after reversing the
flow direction (i.e. when droplets flow from narrow channels into
widened sections), the queues are suppressed -- resulting in uniform
distribution of droplets in the loop. 

Our results can be used for further studies on the dynamics of flow in
microfluidic networks. They can be also directly applied for
randomizing flow of droplets through branches (with channels that widen
in the downstream direction), for alternate directioning long queues to
one of two outputs (with narrowing channels), and for simple, low-cost
generation of periodically changing difference of flow between
branches. The latter may be used for example for scanning the processes
of splitting or collisions of droplets in a T-junction as a function of
the difference of flow rates.

The results described here may be also
helpful in understanding the flow of blood in branched systems of
vascular capillaries.
Interestingly, the smallest capillaries are equipped with a mechanism of
regulated constrictions at the initial section of the capillary:
the regulation of lumen is provided by smooth muscles called
precapillary sphincters, or possibly by other mechanisms called
precapillary resistance\cite{rhoades2009medical}.
Could these constrictions be the evolutionarily developed mechanism of
suppressing oscillations and promoting uniform distribution of red blood cells?
In fact, giant oscillations were recently found in artificial microvascular
networks\cite{ForouzanBloodOsc} and earlier in theoretical models of
blood capillary network\cite{carr2005osc}, although they were never
observed in real microvascular networks.

Oscillations similar to our finding were recently discovered in
a stratified flow of two \emph{miscible} liquids in a narrowing
loop\cite{GeddesWaterGlycerolOsc}.
It suggests a possible existence of a common mechanism of oscillations
caused by flow of complex fluids through parallel, narrowing ducts.

\begin{acknowledgments}
This project was operated within the Foundation for Polish Science Team
Programme co-financed by the EU European Regional Development Fund and
within the European Research Council Starting Grant 279647. S.J. and
O.C. acknowledge financial support from the Polish Ministry of Science
under the grant Iuventus Plus nr IP2012 015172. The authors thank Jakub
Checinski, Tomasz Smolka, Szymon Bacher and Michal Dabrowski for
valuable help during their student practices.
\end{acknowledgments}

\bibliography{bibtex}
\end{document}